\documentclass[review]{elsarticle}

\usepackage{amsmath}
\usepackage{amsfonts}
\usepackage{multirow}
\usepackage{graphicx,psfrag,epsf}
\usepackage{enumerate}
\usepackage{natbib}
\usepackage{url} 

\usepackage{lineno,hyperref}
\modulolinenumbers[5]

\journal{Mathematics and Computers in Simulation}

\begin{document}

\begin{frontmatter}

\title{Stochastic Modelling of Random Access Memories Reset Transitions}

\author[mymainaddress]{M. Carmen Aguilera-Morillo}
\address[mymainaddress]{Department of Statistics. University Carlos III of Madrid, Spain}

\author[mysecondaryaddress]{Ana M. Aguilera\corref{mycorrespondingauthor}}
\cortext[mycorrespondingauthor]{Corresponding author}
\ead{aaguiler@ugr.es}
\address[mysecondaryaddress]{Department of Statistics and O.R. and IEMath-GR, University of Granada, Spain}

\author[mytertiaryaddress]{Francisco Jim\'enez-Molinos}
\address[mytertiaryaddress]{Department of Electronics and Computer Technology. University of Granada, Spain}

\author[mytertiaryaddress]{Juan B. Rold\'an}

\begin{abstract}
Resistive Random Access Memories (RRAMs) are being studied by the industry and academia because it is widely accepted that they are promising candidates for the
next generation of high density nonvolatile memories.
Taking into account the stochastic nature
of mechanisms behind resistive switching, a new technique based on the use of functional data analysis has been developed to accurately model
resistive memory device characteristics. Functional principal component analysis (FPCA) based on Karhunen-Lo\`eve expansion is applied to obtain
an orthogonal decomposition of the reset process in terms of uncorrelated scalar random variables. Then, the device current has been accurately
described making use of just one variable presenting a modeling approach that can be very attractive from the circuit simulation viewpoint.
The new method allows a comprehensive description of the stochastic variability of these devices by introducing a
probability distribution that allows the simulation of the main parameter that is employed for the model implementation.
A rigorous description of the mathematical theory behind the technique is given and its application for a broad set of experimental measurements is explained.
\end{abstract}


\end{frontmatter}


\section{Introduction}

Since the presentation of Moore's law fifty years ago \cite{moore}, the scaling of electronic devices has been increasing continuously till now. Gordon Moore observed that the number of components within integrated circuits double approximately every 18 months. Since that time, in the sixties, the production lines in the electronic industry have been working under pressure with the objective of fulfilling Moore's law. The most important internal computer circuits: microprocessors and memories also have fallen under the influence of the scaling trend imposed by this law \cite{moore, xie2014emerging}.

The reduction of semiconductor devices is not an easy task, the path of scaling is flooded with physical and technological hurdles that sometimes can not be solved. This situation is found in different facets of the integrated circuit industry, in particular in the non-volatile memory realm. The basic components found in non-volatile memory chips, floating-gate transistors, are thought to be facing important limits \cite{xie2014emerging}; therefore, other emerging technologies are under study both in the industry and the academia. Among them there can be found resistive switching (RS) memories \cite{xie2014emerging, waser2007nanoionics, pan, lee}, phase change memories (PCM) \cite{pcm} and Spin-transfer torque random access memories (STT-RAM) \cite{sst}.

RRAMs can store information without the need of a power source when they are switched off. These devices, with an operation based on resistive switching mechanisms, show interesting characteristics such as fast switching speed, endurance, low power operation and compatibility with current complementary metal-oxide-semiconductor CMOS technology \cite{waser2007nanoionics, strukov2008missing}. Because all this, they are considered the most promising future technology for non-volatile memories \cite{xie2014emerging, lee, pan}.

Each generation of devices developed in the microelectronic industry needs to be completely characterized; i.e., the electric currents, capacitances and other magnitudes have to be measured and modeled in such a way that they can be calculated versus the voltages applied at their terminals by means of analytical equations. These analytical expressions are employed in circuit simulators, using also Kirchhoff's laws, to design electronic circuits. A set of equations and the corresponding fitting parameters that characterize an electron device for circuit simulation is known as a compact model and they are under continuous development due to new physical effects that show up as device dimensions
are reduced and new materials and technological processes are employed in their fabrication.

A compact model is essential for the introduction of a technology since the fabrication of new integrated circuits is extremely complicated if previous circuit simulations can not be performed. Although several authors have published models for RRAMs \cite{rpicosmemrysis2015,biolek2009spice, jimenez2015spice, shin2010compact}, there is a long way to go in this field, mostly taking into account the variety of physical mechanisms employed to explain the physics behind the operation of dozens of different RRAMs to date \cite{waser2007nanoionics}. It is well known that the mechanisms behind resistive switching, the core of RRAM operation, are stochastic \cite{pan, lee}. The formation of
conductive filaments that help to drastically change the resistance of the device, from a High Resistance State (HRS) to a Low Resistance State (LRS) is known as a set process, the reverse transition is known as a reset process; in this latter case, the conductive filament is destroyed. These filaments are formed by the random clustering of metallic ions or oxygen vacancies \cite{pan, lee}. That is why a mathematical model and an analysis tool that allow the description of the current-voltage curves by considering the stochastic nature of the device operation are highly desirable. The present study addresses this problem by proposing a novel approach based on Functional Data Analysis (FDA) methodologies for modeling and reconstruction of RRAM current-voltage curves. As far as we know, it is the first time that FDA is used to model these type of reset/set stochastic processes. The interested reader is referred to \cite{RamsayI, RamsayII, FerratyVieu, RamsayIII, Horvath} for a detailed study of the theoretical, computational and applied aspects of the most basic FDA statistical methods.

FDA is a very successful subject of statistical research where the data units are functions of a continuous domain instead of vectors as in classical multivariate analysis. In the most usual case, the functions are curves defined on a real interval of time or other continuous magnitude as voltage in the case of RRAM reset/set processes. The dimension reduction technique Functional Principal Component Analysis (FPCA) is performed in this paper to provide an approximated orthogonal decomposition of the stochastic process generating the current-voltage curves in terms of a finite set of uncorrelated scalar random variables that explain the main features of process variability \cite{Deville1974}. FPCA is based on the well known Karhunen-Lo\`eve expansion introduced by \cite{Loeve1946, Karhunen1946}. FDA techniques  have  been successfully applied in chemical, physics, engineering  and mathematics, including many other interdisciplinary areas
\cite{Fdez2006, Aguilera2010, Hall2012, Zhou2013}
\linebreak \cite{Lillo2016, Delicado2017, Menafoglio2018, Portela2018}.

In our analysis, experimental measurements from a sample of reset cycles have been employed to formulate and estimate the statistical model in a comprehensive and coherent manner. In FDA, the first step, previous to apply a concrete methodology, is to reconstruct the mathematical shape of the curves over their entire domain. The problem related with having discrete observations of the reset process with different domains among sample curves (their domains are upper bounded by the reset voltage) is solved by a new approach based on synchronization of the observed curves in a common interval. Then,  FPCA is estimated by using P-spline smoothing in terms of basis representation with B-spline functions \cite{AguileraSPCA}.  In this way some of the well-known FDA features are employed here to their full extent in order to capture the random characteristics of RRAM operation. The most important goal in this paper is to study and model important  patterns of variability among the data (the goals of functional data analysis are essentially the same as those of any other branch of statistics \cite{RamsayI}, Chap. 1).

The manuscript is organized as follows. In Section~2, we describe the technological details of the fabricated devices and observational data. In Section~3 the features of the modeling procedure based on FPCA are explained. In Section~4 the main results and the corresponding discussion are given. Finally, in Section~5 we wrap up the contents presented along the paper.

\section{Device fabrication and observational data}

The  devices studied were fabricated at the IMB-CNM (CSIC) in Barcelona. They are based on a Ni/Hf$O_{2}/Si-n^{+}$ structure, the dielectric layer was 20nm thick; other details of the fabrication process and measurement setup are given in \cite{gonzalez2014analysis}. In these memories, conduction takes place inside conductive filaments (CFs) that are formed and destroyed within the RS device operation \cite{gonzalez2014analysis, villena2013depth}.

In this work,  only experimental measurements are used. Nevertheless,  simulated results can also be used in future developments since we have implemented a simulation tool that accounts for the main physical effects involved in RRAM operation \cite{villena2014simulation, villena2015depth, Villena2016}.

\begin{figure}[h]
\center 
\includegraphics[width=.6\textwidth]{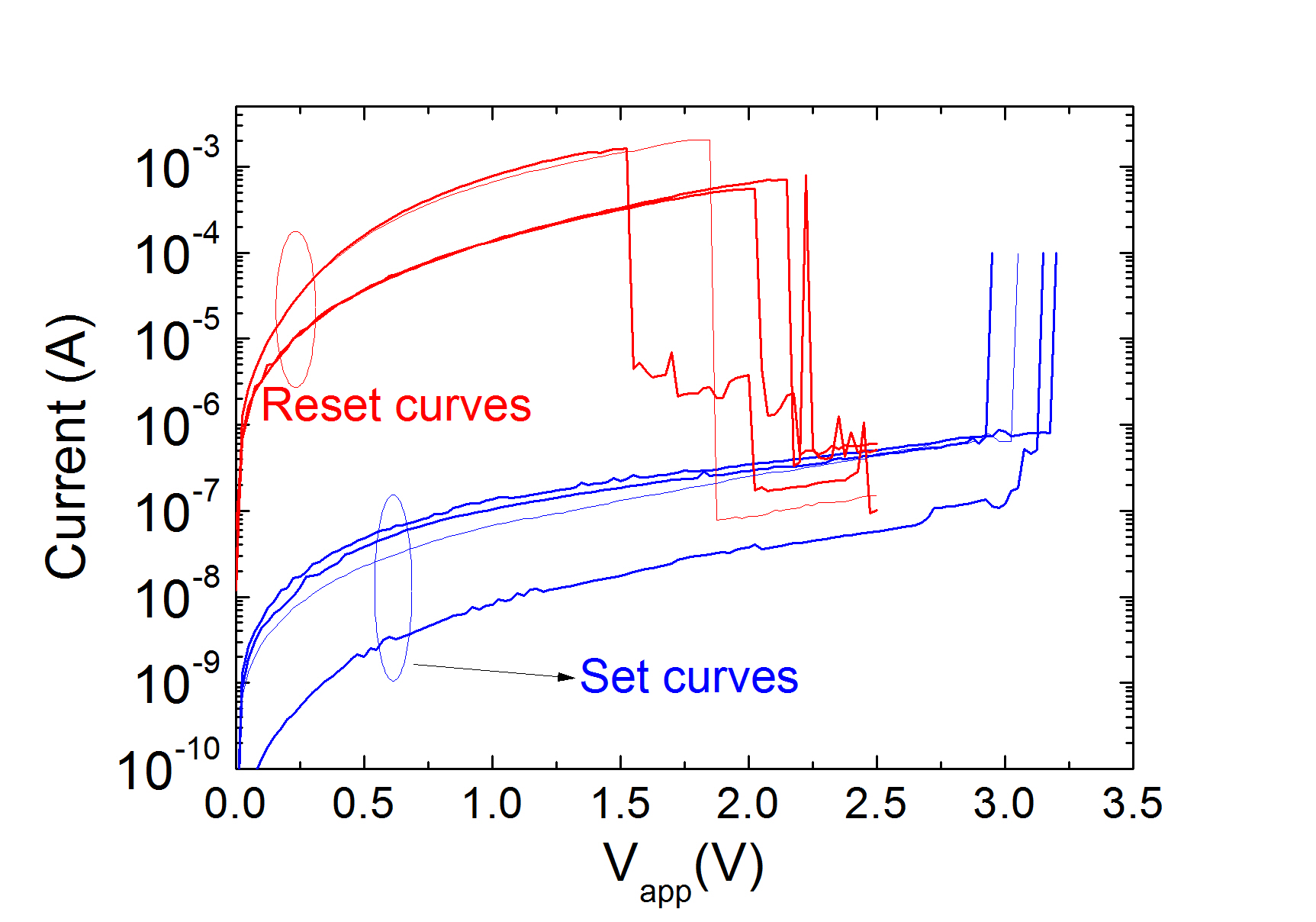}
\caption{Experimental current versus applied voltage for several set/reset transitions in a long set-reset series for devices based on a Ni/Hf$O_{2}/Si-n^{+}$ structure. Although the Ni electrode had a negative voltage applied while the substrate was grounded \cite{gonzalez2014analysis}, we have considered absolute values for the applied voltage in order to ease the modeling process for curves in the first quadrant. The curves have been plotted on a logarithmic scale for the current.}%
\label{fig:figura1}
\end{figure}

A few current-voltage curves corresponding to several set-reset cycles are shown in Figure \ref{fig:figura1}. Reset curves are shown in red lines while set curves are given in blue lines. These curves correspond to different cycles from a set-reset series of three thousands cycles. It can be seen that the curves are different in all the cases, this is due to the stochastic nature of the processes behind the conductive filament formation that determines the device resistance, i.e., the ratio between the device voltage and the corresponding current \cite{xie2014emerging, lee, pan, villena2014simulation}.

\begin{figure}[h]
\center 
\includegraphics[width=.6\textwidth]{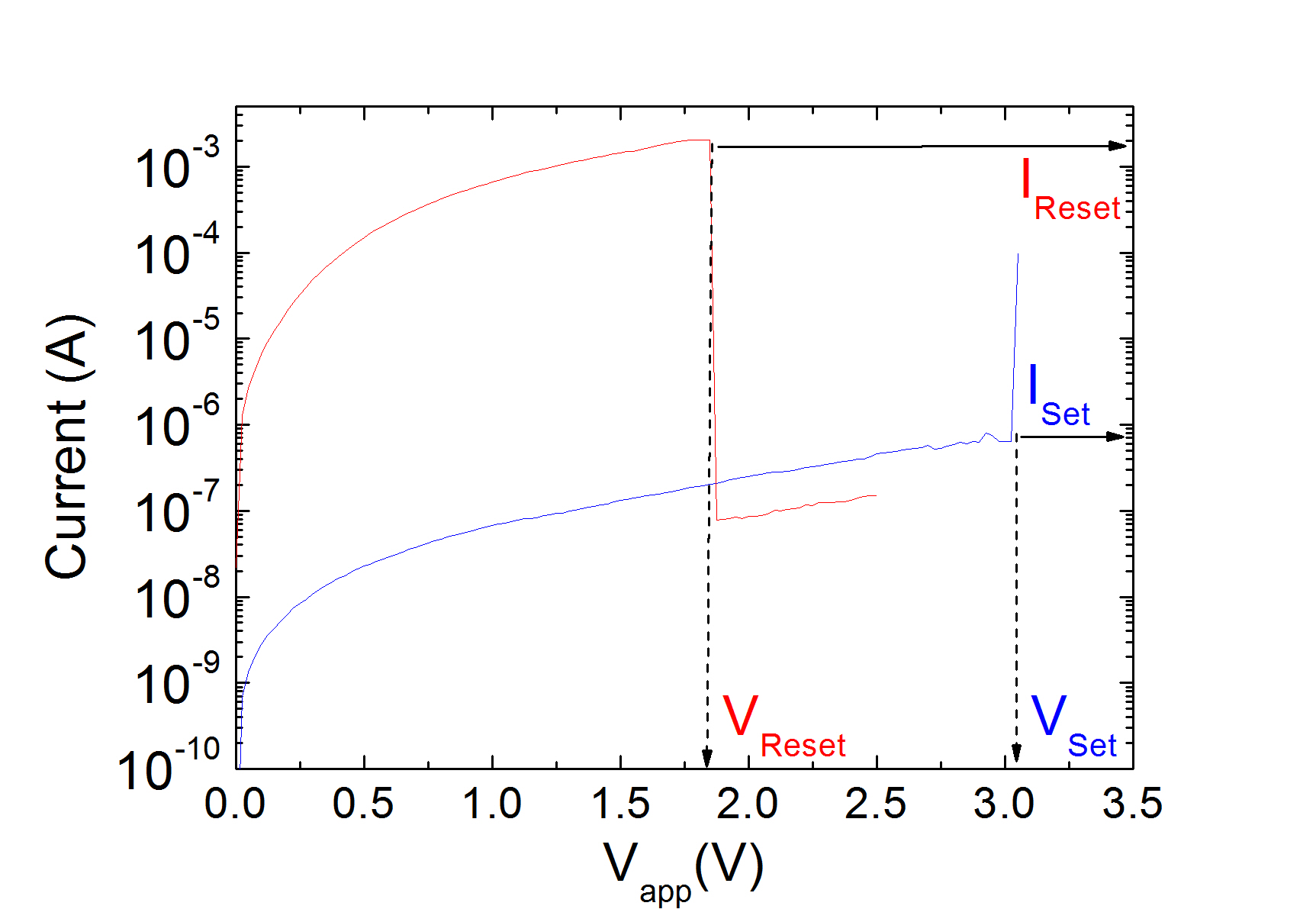}
\caption{Experimental current versus applied voltage for a set and reset transitions for a device based on a Ni/Hf$O_{2}/Si-n^{+}$ structure. The ($V_{Reset}$, $I_{Reset}$) and ($V_{Set}$, $I_{Set}$) points are highlighted.}%
\label{fig:figura2}
\end{figure}

Two of the curves plotted above have been isolated in Figure \ref{fig:figura2}. The $V_{Reset}$ and $I_{Reset}$ points are marked for the reset curve (red line).
For any reset curve, these magnitudes represent the reset voltage and reset current linked to the rupture of the conductive filament \cite{gonzalez2014analysis, villena2013depth}. So the reset  point is determined by the sudden drop of the current and at this point the device enters the high resistance state because the resistance of the conductive filament is much lower than the dielectric resistance. The $V_{Set}$ and $I_{Set}$ points are also marked for the set curve (blue line). For any set curve, these magnitudes mark the set voltage and set current where the creation of the conductive filament takes place. The filament formation makes the RRAM resistance greatly diminish because of the low resistance of the conductive filament.

As it will be shown in the next section, the reset voltage  is a key parameter for our modeling procedure, since all the reset curves studied are defined between zero and the corresponding $V_{Reset}$, which is different among them. Because of this, all the curves are first  normalized in the interval [0,1]. Then,  the other FDA steps are performed.

\section{Functional data analysis}

As stated in the introduction, Functional Data Analysis is performed in this paper for modeling and explaining the stochastic variability of  the curves  of evolution of electric current (I) in terms of voltage (V) from a sample of reset cycles in Resistive Random Access Memories (RRAMs).  Let us denote by $\{ I (v):  v\in T \}$ the stochastic process of evolution of current in terms of voltage in a real interval T.

In order to reduce the dimension  and  explain the main features and modes of variation of the current-voltage reset process, the present study perform a new approach based on functional principal component analysis (FPCA) that was introduced by  \cite{Deville1973,Deville1974} as a generalization of  multivariate principal component analysis to
the case of a continuous-time stochastic processes. FPCA is  based on the well known Karhunen-Lo\`eve expansion (KLE) that makes and orthogonal decomposition of the process in terms of uncorrelated random variables and deterministic functions. This method was first introduced   as harmonic analysis of a stochastic process. Asymptotic theory and statistical inference on FPCA were developed in \cite{Dauxois1982}.

\subsection{Karhunen-Lo\`eve expansion}

Let us suppose that
$\{I (v) \}$
is a second order stochastic process defined on a probabilistic space $(\Omega,A,P),$ continuous in  quadratic  mean
and whose sample functions
belong  to the Hilbert space L$^2[T]$ of the square
integrable  functions
on $T$, with the natural inner product defined by
$$
<f|g> := \int_T f(v)g(v)dv \ \  \mbox{for all} \ \ f,g\in L^2[T].
$$
     The  covariance  operator  $\cal C$  of  $\{I(v) \}$ is a positive
autoadjoint  compact  operator defined on L$^2 [T]$ by
$$
{\cal C} (f(v)) := \int_T C(v,u) f(u) ds,
$$
with kernel  the   covariance function $C(v,u)$,
     Then, the spectral representation of $\cal C$ provides the following
orthogonal decomposition of the process, known as  Karhunen-Lo\`eve
orthogonal expansion \cite{Karhunen1946,Loeve1946}:
\begin{equation}
I(v)  = \mu (v) + \sum_{j=1}^\infty f_j (v) \xi_j ,
\label{K-L}
\end{equation}
where  $\{f_j \}$  is  the  orthonormal  family  of  eigenfunctions  of
the covariance operator $\cal C$ associated with its decreasing sequence of non null eigenvalues
$\{\lambda_j \}$, that is
$$
{\cal C} (f_j (v))=\int_T C(v,u) f_j (u) du=\lambda_j f_j(v), \quad v
\in T,
\label{eqi}
$$
and $\{\xi_i \}$ is the family of uncorrelated zero-mean  random  variables
defined by
$$
\xi_j :=\int_T f_j (v)(I(v) -\mu(v))dv.
$$
     The random variable $\xi_j$  is called the {\it jth} principal  component  and
has  the maximum  variance  $\lambda_j$ out of all  the  generalized   linear
combinations of $I(v)$  which are uncorrelated with $\xi_k$  $(k=1,..,j-1).$
Similarly, $f_j$  is called the {\it jth}  principal weight function
or harmonic factor.

   Taking into account that the total variance of $\{I(v)\}$ is given by
$$
V:=\int_T C(v,v)dv = \sum_{j=1}^\infty  \lambda_j,
$$
then the ratio $\lambda_j /V$ is the variance explained by the {\it jth}
component.

In addition, the series (\ref{K-L}) truncated  in  the
{\it qth} term is the best approximation of the process (in the least-squares
sense) by a sum of  {\it q}  quasi-deterministic  terms  \cite{Saporta81}.
Therefore the process admits the following principal component reconstruction:
$$
I(v)^q =\mu(v)+\sum_{j=1}^q f_j(v) \xi_j ,
\label{rec}
$$
in terms of the first {\it q}
principal components so that the sum of the variances explained
by  them  is  as  close  as  possible to one.

\subsection{Sample FPCA}

In practice, to estimate the functional principal components and weight functions we have a random sample of functional data that consists of $n$ reset curves denoted by  $$\{ I_i (v): i=1,\dots,n; v\in [0,V_{i-reset}] \},$$
where $V_{i-reset}$ is the voltage to reset. In addition, the data consist of  discrete observations of each reset curve  $I_i (v)$  at a finite set of current values until the $V_{i-reset}$ point. Concretely, each curve  $I_i (v)$  is observed at  $k_i = V_{i-reset}*10^3$ discrete equally spaced  sampling points  $v_j = j*10^{-3}$ $(j=1,\dots,k_i).$

In this context, the estimation of FPCA presents two important problems. On one hand, the  reset curves are  not defined on the same domain because the
voltage to reset  is different among cycles.  On the other hand, we only have discrete observations of each reset curve  at a finite set
of current values until the V$_{reset}$ point.
 To solve these problems we propose a novel FDA approach based on three main steps:
 \begin{enumerate}
 \item Registration of the reset curves in the interval [0,1].
 \item  Reconstruction of the reset curves by P-spline smoothing on the registered data.
 \item Functional PCA    of the basis representation of reset curves in terms of P-splines.
 \end{enumerate}
Let us now briefly summarize each of these steps.

\subsubsection*{Registration}

Let us observe that the reset curves have different domains because the
voltage to reset  is different among cycles.
In this case, the first step in FDA is curve registration (transforming curves by transforming their arguments (see \cite{RamsayI}, Chap. 7, for a detailed description). In this paper we propose to do it in the simplest way that consists of transforming the domain
 $[0,V_{i-reset}]$ of each reset curve in the interval $ [0,1]$ by the function $v/V_{i-reset}.$ Then, FDA methodologies will be performed on the synchronized curves given by
$$I^*_i (u) := I_i (u*V_{i-reset}) \quad \forall u\in [0,1],$$ so that  $$I_i (v) = I_i^* (v/V_{i-reset}) \quad \forall v\in [0,V_{i-reset}].$$

This way, for each curve we have a new set of arguments in the interval [0,1] given by  $$u_{ij} := \frac{v_j}{V_{i-reset}} = \frac{j}{k_i} \ \ (j=1,\dots,k_i).$$ This means that the sampling points where the registered curves are observed are different not only in number but also in position.

\subsubsection*{Basis representation and smoothing}

In order to reconstruct the true functional form of registered reset curves we will assume that they belong to a
finite-dimension space spanned by a basis $\left\{ \phi_{1}\left(
t\right) ,\ldots,\phi_{p}\left(t\right)  \right\}$, so that they are
expressed as
\begin{equation}
I_{i}^* \left(  u\right)  =\sum_{j=1}^{p} a_{ij}\phi_{j}\left( u\right)
,\;i=1,\ldots,n.
\label{representa}
\end{equation}
The selection of the  basis and its dimension $p$  is crucial
 and must be done according to the characteristics of the curves. Useful basis systems are Fourier basis for periodic
data, B-spline basis for non-periodic smooth data with continuous derivatives up to certain order, and wavelet basis for data
with a strong local behavior whose derivatives are not required \cite{RamsayII, Aguilera1996, Aguilera2014}.
Assuming that the reset curves are smooth and  observed
with error
$$I_{ij}^* = I_i^* (u_{ij}) + \epsilon_{ij} \quad j=0,1,\dots, k_i,
i=1,\dots,n,$$
least squares approximation with B-splines basis is an
appropriate choice to approximate the basis coefficients $a_{ij}.$

A B-spline basis of order $p+1$ (degree $p$) generates the space of the splines of the same
degree, defined as curves consisting of piecewise polynomials of degree $p$ that
join up smoothly at a set of definition knots with continuity in their derivatives
up to order $p-1.$ A detailed study of these bases can be seen in \cite{DeBoor2001}. A pioneer work on data analysis with splines was developed by \cite{Wold1974}.

In this paper,  the  iterative definition of B-splines
introduced by \cite{DeBoor1977} is considered.
Denoting the definition knots by $t_{0}<\ldots <t_{m},$ and extending this partition of the domain as
$t_{-p}< \ldots < t_{-2}<t_{-1}<t_{0}<\ldots <t_{m}<t_{m+1}<t_{m+2}< \ldots < t_{m+p}$,
the basis of B-splines of order $p+1$ (degree $p)$ is iteratively defined by
\begin{eqnarray}
B_{j,p+1}\left( t\right) &:=&\displaystyle\frac{t-t_{j-2}}{t_{j+p-2}-t_{j-2}}%
B_{j,p}\left( t\right) +\displaystyle\frac{t_{j+p-1}-t}{t_{j+p-1}-t_{j-1}}%
B_{j+1,p}\left( t\right) \nonumber \\
&&  \nonumber \\
p &=&1,2,\ldots ;\;j=-1,0,\ldots ,m-p+4,  \nonumber
\end{eqnarray}
with
$$
B_{j,1}\left( t\right) :=\left\{
\begin{tabular}{lll}
$1$ &  & $t_{j-2}\leq t<t_{j-1}$ \\
$0$ &  & otherwise
\end{tabular}
\right. ,\;j=-1,0,1,\ldots ,m+4.
$$

 The curves fitted by ordinary least squares approximation
in terms of B-spline basis are known as
regression splines and their main problem is that they don't control the
degree of smoothness. Penalized spline smoothers are usually considered to solve this problem  by introducing a penalized least squares approach that measures the roughness of the curves. Smoothing splines (continuous roughness penalty) were applied to improve the extraction of threshold voltage in MOSFETs transistors  (\cite{Ibanez2014, Ibanez2015}). In this  paper, P-splines  (discrete roughness penalty) are used for reconstructing the reset curves.  They  measure the roughness of the curves  by summing squared d-order differences between
adjacent B-splines and their  main advantage is that the
number of knots is not so determinant as in regression splines and can be
easily compute (\cite{Eilers199689, Eilers2015}). As the reset curves are smooth enough  to ensure an accurate spline approximation, an appropriate alternative to P-splines could be  using regression splines and choosing the dimension of the B-spline basis by cross-validation.

For each reset curve, the basis coefficients of the penalized spline smoother in terms of
B-spline basis functions are  computed by minimizing the penalized
least squares error
$$
PMSE_d \left(a_{i}|I_{i}^*
\right):=\left(I_{i}^*-\Phi_{i}a_{i}\right)'\left(I_{i}^*-\Phi_{i}a_{i}\right)+\lambda
a_{i}'P_{d}a_{i}, \label{minCuadPenPspl}
$$
with $I_{i}^*=\left(I_{i1},I_{i2},\ldots,I_{ik_{i}}\right)^{\prime}$ being the vector of discrete measures of the registered curve $I_i^*(u),$
$\Phi_{i}:=\left(\phi_{j}\left( u_{ij}\right)
\right)_{k_{i}\times p}$ being the matrix of values of the basis functions at the sampling points and
$ P_{d}:=\left(\triangle^{d}\right)'\triangle^{d}$ with
$\triangle^{d}$ being the matrix representation of the d-order
difference operator.

Then, the
B-spline basis coefficients for each curve are given by
$$
\hat a_{i}=\left(\Phi_{i}'\Phi_{i}+\lambda
P_{d}\right)^{-1}\Phi_{i}'I_{i}^*.
$$

In general, the knots
of a P-spline must be equally spaced and its number sufficiently large to fit
the data and not so large that computation time is unnecessarily big.
There are some important
choices related to the P-spline fitting: the smoothing parameter,
the order of the penalty, the degree of the B-spline basis  and the
number of knots. The simplest and most usual choice for these  parameters that should work well in most applications is
using cross-validation for chosen the smoothing parameter, a quadratic penalty, cubic splines and one knot for every four
or five observations up to a maximum of about $40$ knots (see \cite{AguileraPsplines} for a comparative study of the  performance of regression splines, smoothing splines and P-splines on simulated and real data).

In order to select the same smoothing parameter for all the n sample
paths we propose to minimize the mean of the leave-one-out cross validation errors
over all sample curves.

The leave-one-out  cross validation  (CV) method consist of selecting the smoothing parameter
$\lambda$ that minimizes
$$
CV (\lambda) =\frac{1}{n}\sum_{i=1}^{n} CV_i (\lambda),
$$
where $$CV_i (\lambda) =  \sqrt{
\sum_{j=0}^{k_{i}}\left(I^*_{ij}-\hat{I^*}_{ij}^{-j)}\right)^{2}/(k_i+1)},
$$
with $\hat{I^*}_{ij}^{-j)}$ being the values of the \emph{i}-th sample
path estimated at time $t_{ij}$ avoiding the \emph{j}-th time point in
the iterative estimation process.

A computationally simplest approach very used  in
the literature about smoothing splines is generalized cross-validation (GCV) \cite{Craven1978377}. The
 GCV method consist of selecting $\lambda$ which minimizes
$$
GCV \left(\lambda \right) =\frac{1}{n} \sum_{i=1}^{n} GCV_{i} \left(\lambda \right),
$$
where
$$
GCV_{i} \left(\lambda \right)
=\frac{{(k_{i}+1)} MSE_{i}  \left(\lambda \right) }{[trace \left(I-H_{i}  \left(\lambda \right) \right)]^{2}},
$$
with
$MSE_i  \left(\lambda \right) =\frac{1}{n}\sum_{j=0}^{k_{i}} \left ( I^*_{ij}- \hat{I^*}_{ij}
\right)^{2},$ $H_{i}  \left(\lambda \right) =\Phi_{i}\left(\Phi_{i}'\Phi_{i}+\lambda
P_d\right)^{-1}\Phi_{i}',$ $\Phi_{i}$ being the B-spline basis evaluated at the observation knots and
$P_d$ the discrete penalty matrix.

As alternative to the methodology exposed above, the relation between P-splines and BLUP (best linear and unbiased predictor) in a mixed model allows, in some cases, to use the existing methodology in the field of mixed models in order to select a common smoothing parameter for all curves. In fact, this idea is based on writing a non-parametric or semi-parametric model as a mixed model \cite{Brum98, Verbyla99}. Using the mixed model framework it is possible to estimate the smoothing parameter together with the
rest of the parameters of the model, instead of using cross validation algorithms. The standard method for the estimation of variance components in
mixed models is the restricted maximum likelihood method (REML).

\subsubsection*{Sample estimation}

The sample estimate of the j-th principal component score associated with the registered reset process is  given by
$$
 \hat{\xi}_{ij}^*:=\int_{0}^1 \left ( I_{i}^* \left(u\right) - \bar{I}^* (u) \right ) \hat{f}_{j}^* \left(u\right)du
,\;i=1,\ldots,n, \label{FCP}
$$
where $\bar{I}^* (u)$ is the sample mean function
$$
\bar{I}^* (u) := \frac{1}{n} \sum_{i=1}^n I_i^* (u),
$$
and the weight functions $ \hat{f}_j^*$ are the eigenfunctions of
the sample covariance operator $ \hat{\cal C}^*.$ That is, the solutions to the
second order eigenequation
$$
\hat{\cal C}^* (f_{j}^*) (u) = \int_0^1  \hat{C}^* \left(u,v\right)  \hat{f}_j^* \left(v\right)dv =
\hat{\lambda}_{j}^* \hat{f}_j^* (u), \label{ie}
$$
where $ \hat{C}^* \left(u,v\right)$ is the sample covariance function
$$
 \hat{C}^*(u,v) := \frac{1}{n-1} \sum_{i=1}^n (I_i^* (u)- \bar{I}^* (u))(I_i^* (v)- \bar{I}^* (v)).
$$
Then, an approximation of the sample curves is computed by truncating the KLE in terms of the first $q$ principal components
$$
I_{i}^{*^{q}} \left(u\right)= \bar{I}^* (u)+\sum_{j=1}^{q} \hat{\xi}_{ij}^* \hat{f}_j^* \left(u\right), $$
whose explained variance is given by $\sum_{j=1}^q \hat{\lambda}_j^*.$

Let us  consider that the registered sample paths are expressed in terms of
basis functions as in \ref{representa} and let us denote by $A=\left(a_{ij}\right)_{n \times p}$
the matrix of basis coefficients. Then, the
principal component weight function $\hat{f}_j^*$ admits the basis expansion
$$
\hat{f}_j^* \left(u\right)=\sum_{k=1}^{p}b_{jk} \phi_k\left(u \right),
$$
and FPCA is equivalent to  multivariate PCA of matrix
$A\Psi^{\frac{1}{2}}$ \cite{Ocana2007}, with $\Psi^{\frac{1}{2}}$ being the squared
root of the matrix of inner products between basis functions $\Psi :=\left(\Psi_{ij}\right)_{p \times p} :=\int_T \phi_i
\left(u\right) \phi_j \left(u\right) du$.

Then, the vector $b_j$ of basis coefficients  of the j-th principal
weight function is given by $b_j = \Psi^{-\frac{1}{2}}u_j,$ where
the vectors $u_j$ are computed as the solutions to the eigenvalue
problem
$
n^{-1}\Psi^{\frac{1}{2}}A'A\Psi^{\frac{1}{2}}u_j=\lambda_j^* u_j,
$
where $n^{-1}\Psi^{\frac{1}{2}}A'A\Psi^{\frac{1}{2}}$ is the sample
covariance matrix of $A\Psi^{\frac{1}{2}}.$

\begin{table}
\begin{center}
 \begin{tabular}{lr}
 \hline
PC & Percentage of variance \\
\hline
 1 & 97.2723 \\
 2 & 2.3562  \\
 3 & 0.2344 \\
 4 & 0.0989 \\
 \hline
 \end{tabular}
\end{center}
 \caption{Percentages of variance explained by the first four principal components of P-spline smoothing on the registered reset curves.}
\label{table:tabla1}
\end{table}

\section{Results and discussion}

\begin{figure}
\begin{center}
\includegraphics[width=.6\textwidth]{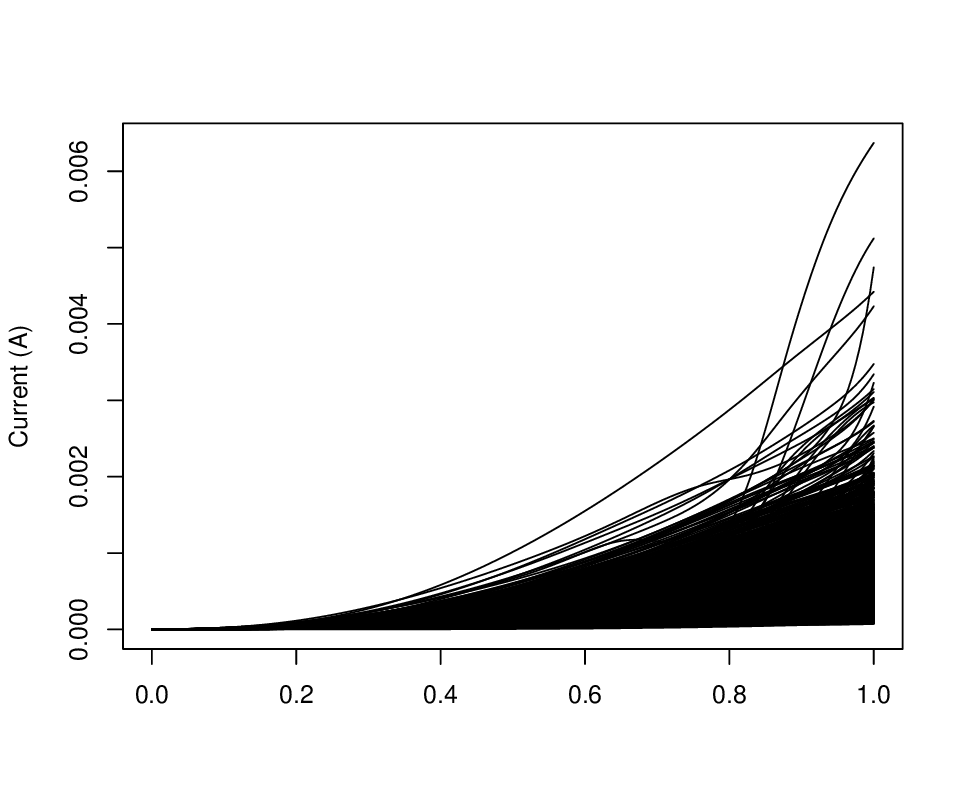}
\caption{Sample of 3057 reset curves obtained for the same device under successive set-reset cycles registered in the interval [0,1].} \label{fig:RegisteredAllTray}
\end{center}
\end{figure}

\begin{figure}
\begin{center}
\begin{tabular}{cc}
\includegraphics[width=.45\textwidth]{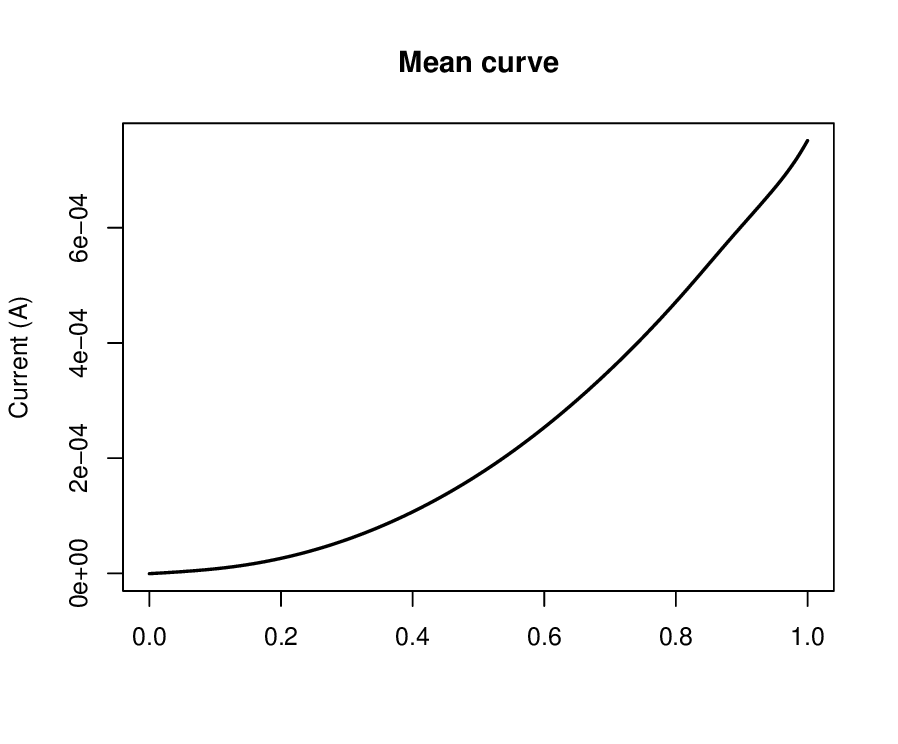}&
\includegraphics[width=.45\textwidth]{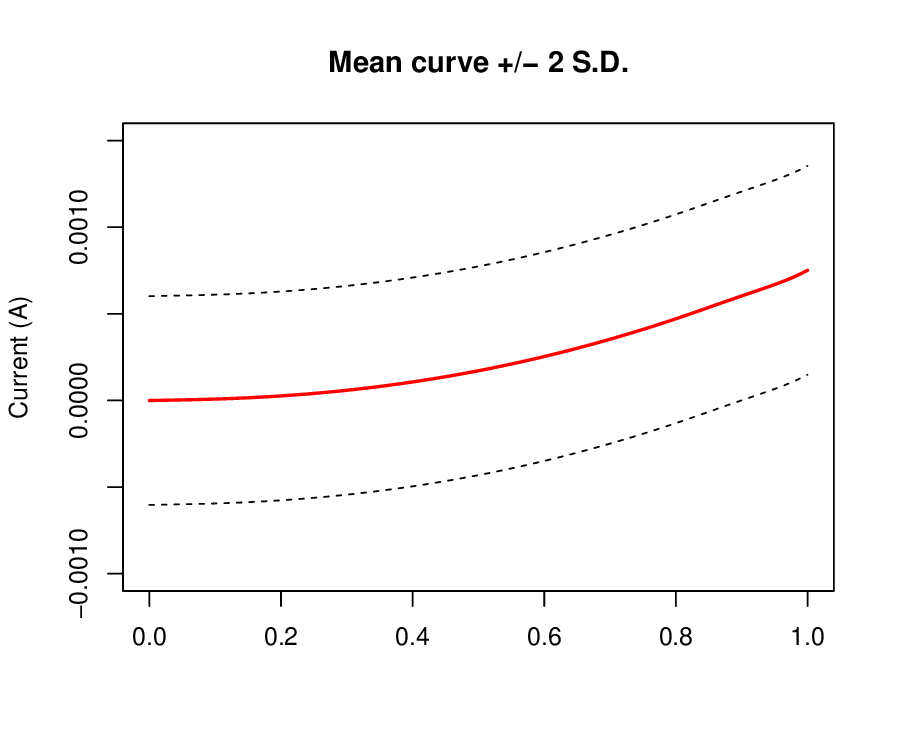} \\
\end{tabular}
\end{center}
\caption{Functional mean of reset curves and confidence bands computed  as $\pm$ 2 times the
standard deviation at each current.} \label{fig:samplemean}
\end{figure}

In order to estimate the principal component decomposition of the reset process previously developed, we have a sample of 3057 reset current-voltage curves obtained for the same device under successive set-reset cycles $$\{ I_i (v): i=1,\dots, 3057 \},$$ observed at the voltage points  $v_j = j*10^{-3}$ $(j=1,\dots)$ until the filament rupture that defines the end of the domain given by the reset voltage $V_{i-reset}.$
FPCA  of the reset process is estimated by following the three steps described in previous section. First, sample reset curves are registered in the interval [0,1]. All the registered curves are displayed in Figure \ref{fig:RegisteredAllTray}. The sample mean function (left) next to pointwise confidence bands are displayed in Figure \ref{fig:samplemean}. Second, P-spline smoothing on the registered discrete data is performed for each of the reset curves in terms of a basis of cubic B-splines defined on 17 equally spaced knots in the interval [0,1]. And third, functional principal components are computed by multivariate PCA on the matrix of basic coefficients appropriately transformed.

\begin{figure}
\begin{center}
\includegraphics[width=.8\textwidth]{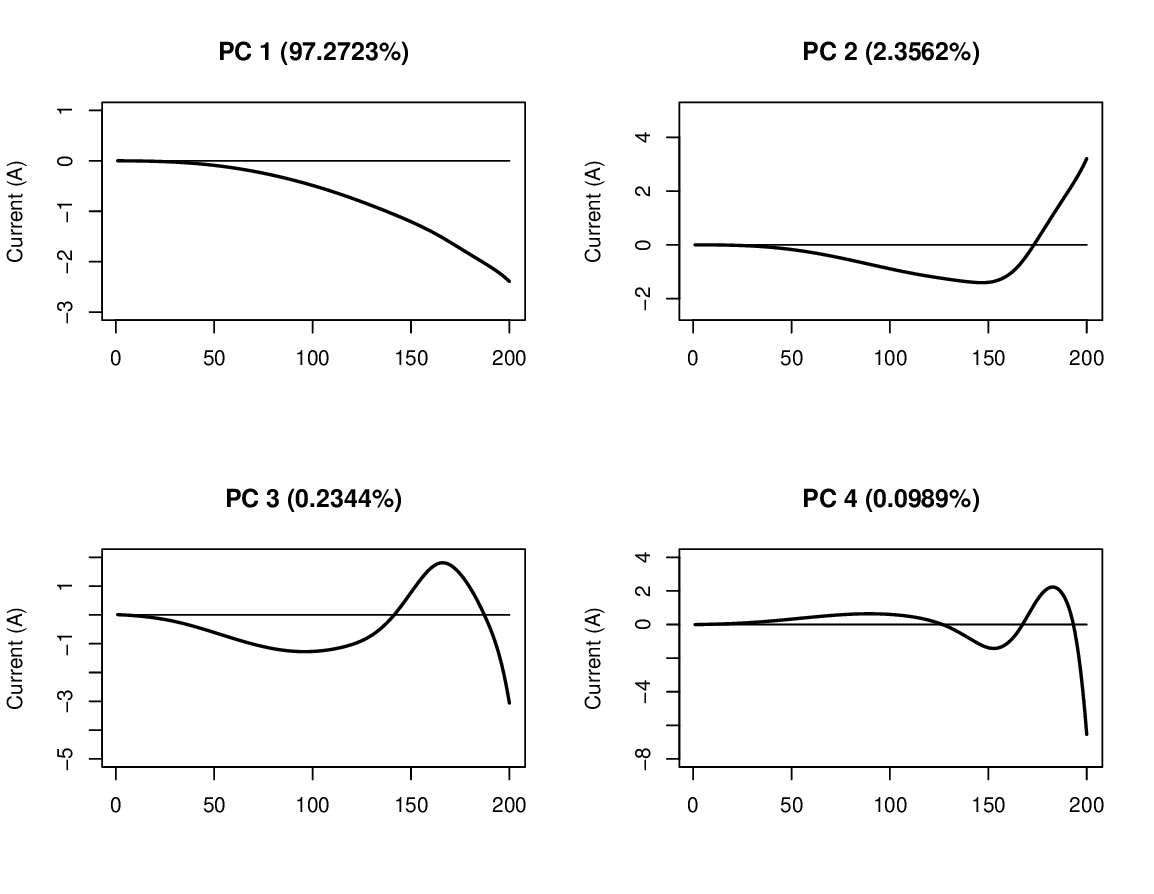}
\caption{First four principal components weigh functions of the registered curves.} \label{fig:PCfactors}
\end{center}
\end{figure}

\begin{center}
\begin{figure}
\begin{center}
\begin{tabular}{cc}
\includegraphics[width=.4\textwidth]{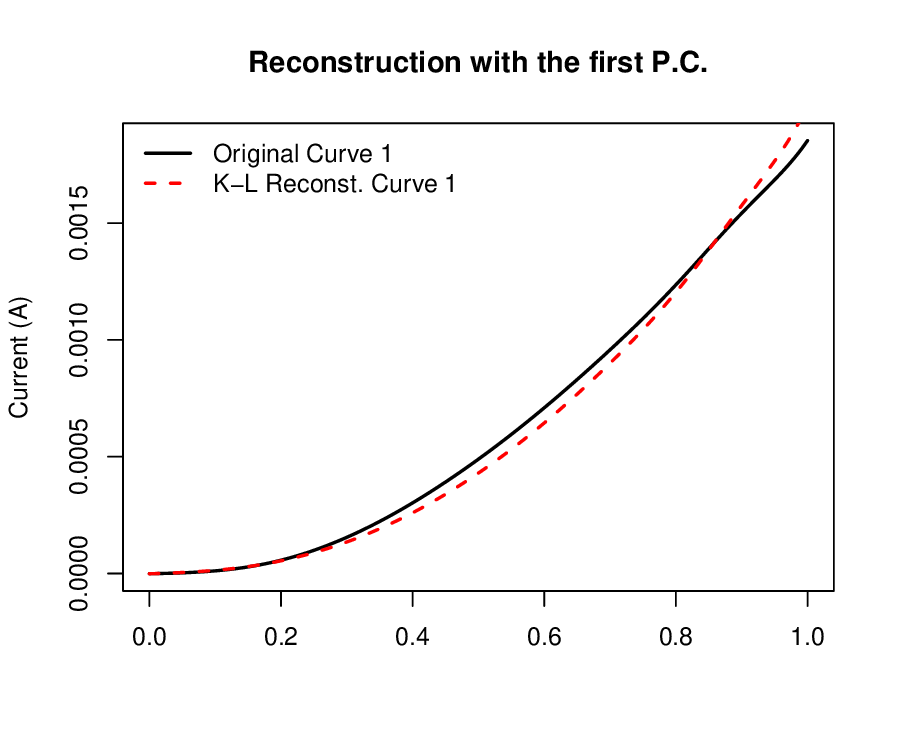}&
\includegraphics[width=.4\textwidth]{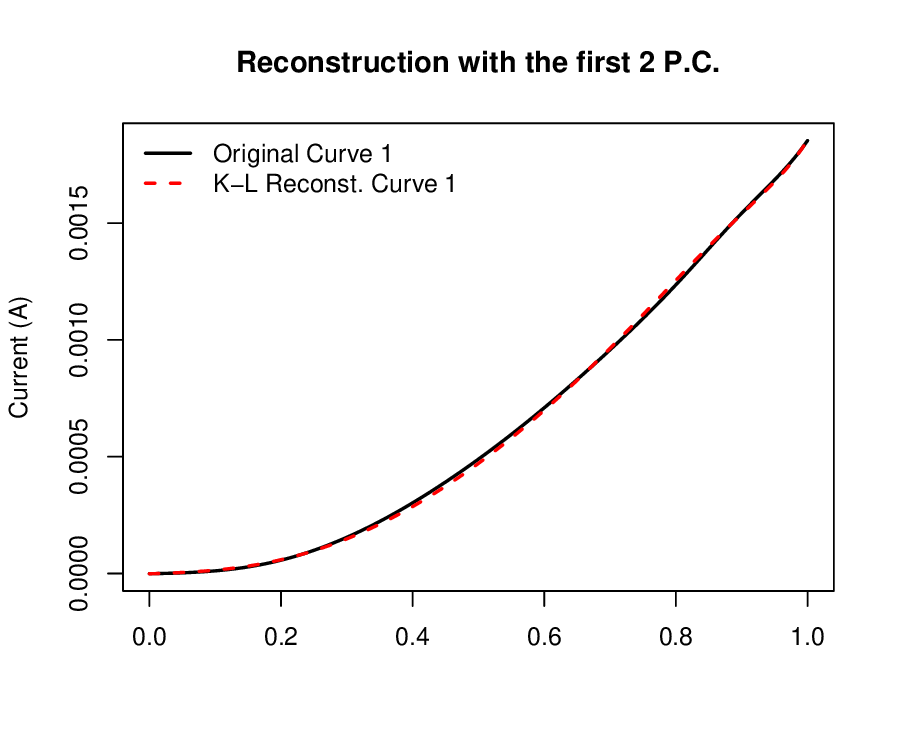}\\
\includegraphics[width=.4\textwidth]{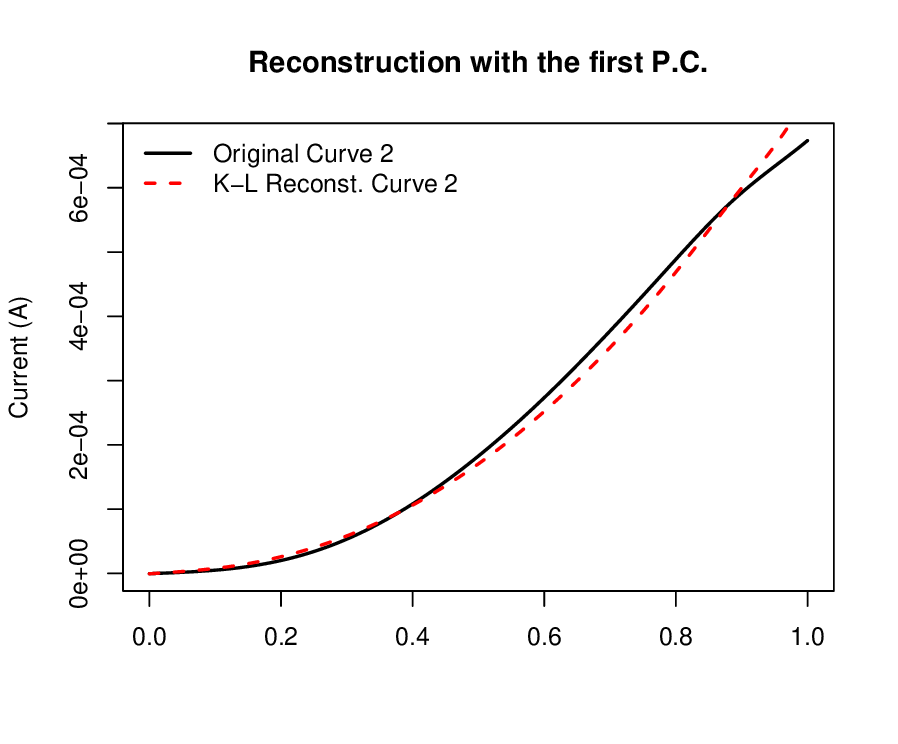}&
\includegraphics[width=.4\textwidth]{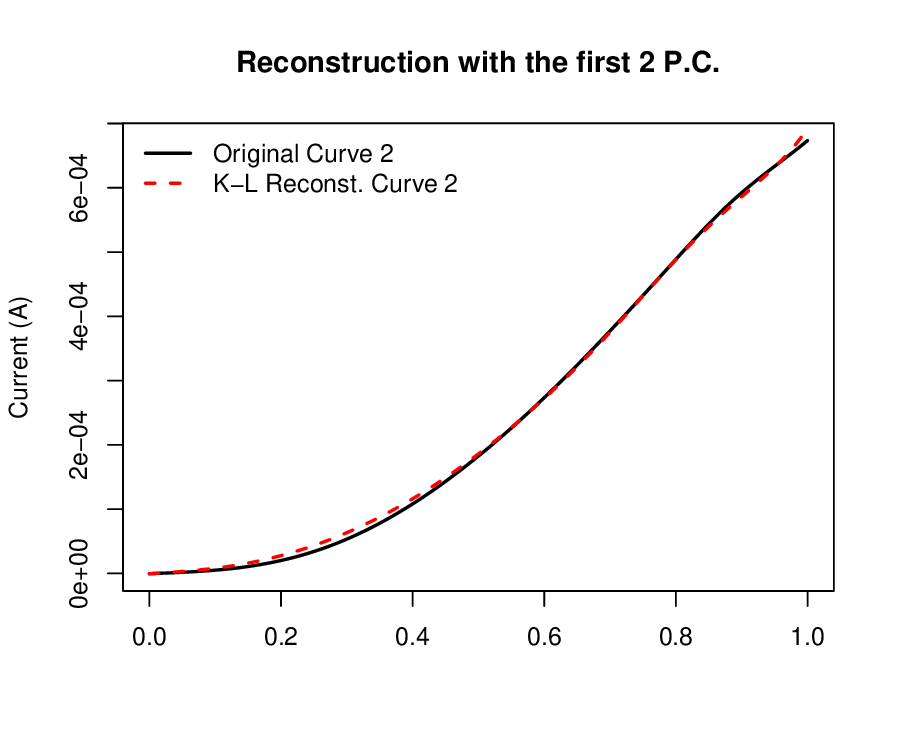}\\
\includegraphics[width=.4\textwidth]{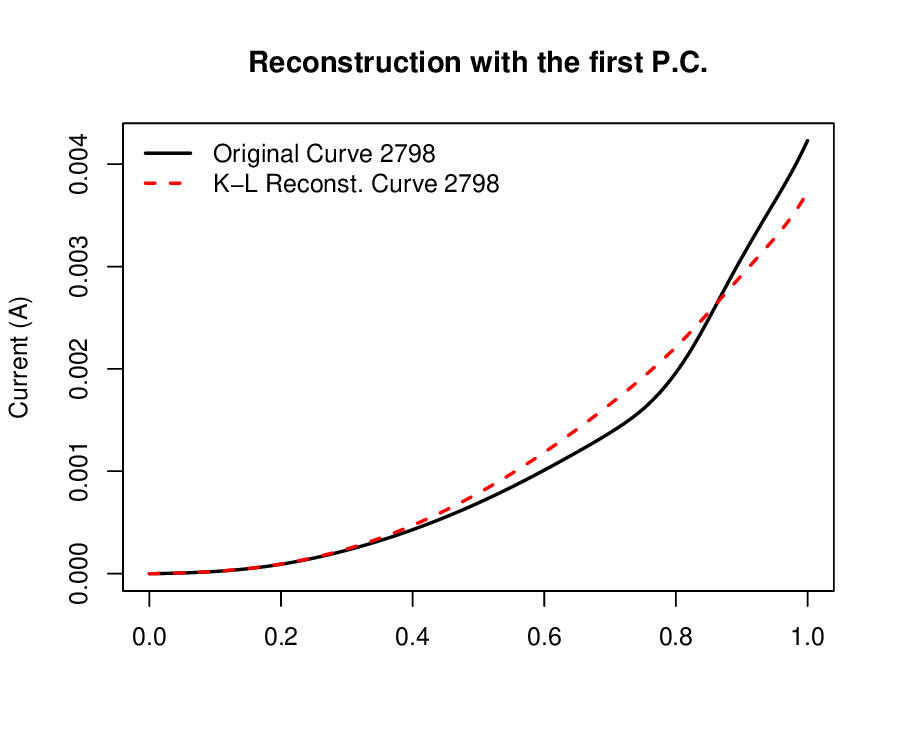}&
\includegraphics[width=.4\textwidth]{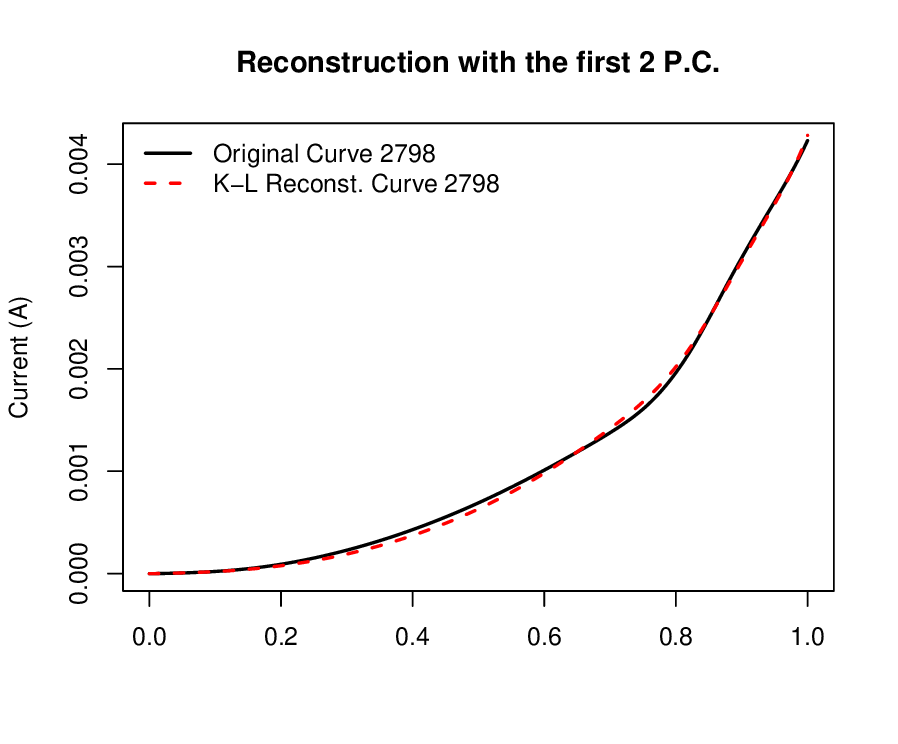}\\
\includegraphics[width=.4\textwidth]{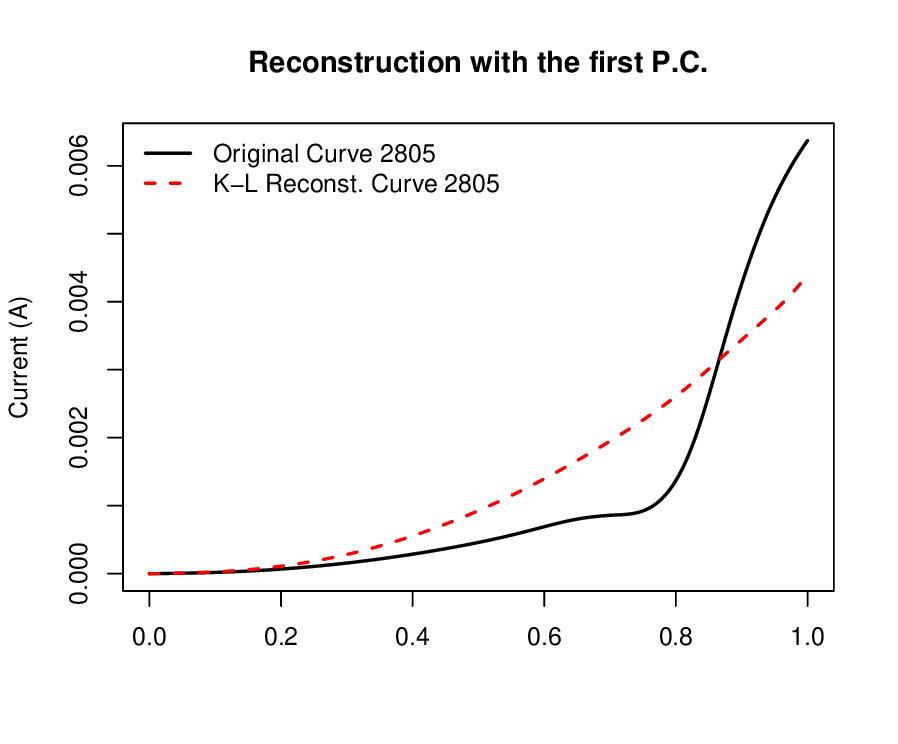}&
\includegraphics[width=.4\textwidth]{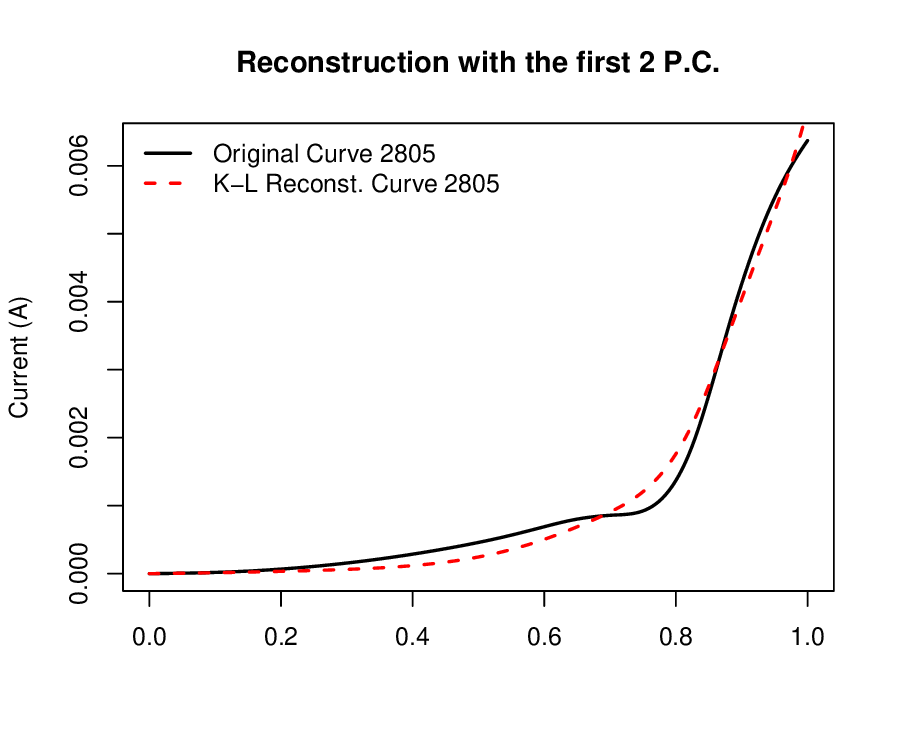}\\
\end{tabular}
\end{center}
\caption{P-spline smoothing of some reset curves (black line) superposed with their
reconstructions (red broken line) in terms of the first p.c. (left)   and the first two pc's (right).} \label{fig:PCreconstruction}
\end{figure}
\end{center}

The principal component weights associated with the first four principal components are shown in Figure \ref{fig:PCfactors}.
The percentages of variance explained by the first four principal components are given in Table \ref{table:tabla1}.
 Let us observe that only the first principal component explains more than a $97\%$ of the total variability of the reset process. Because of this, the principal component decomposition of the registered reset curves can be truncated in the first term providing the following model:
$$
I^{*^1} \left(u\right)= \bar{I}^* (u)+ \xi_{1}^* f_1^* \left(u\right), \ \  u\in[0,1],
$$
where $\xi_{1}^*$ is an scalar random variable (first principal component score) and $f_1^*$ is a deterministic function (principal component weight curve).
The accurated reconstruction given by this principal component decomposition can be seen in Figure \ref{fig:PCreconstruction} where some of the registered curves are approximated in terms of the first principal component. One of the main advantages of this simple linear representation of the reset process is that it could be used for circuit simulation if the distribution of probability of the first principal component is known.

\begin{figure}
\begin{center}
\begin{tabular}{cc}
\includegraphics[width=.5 \textwidth, height =.3 \textheight]{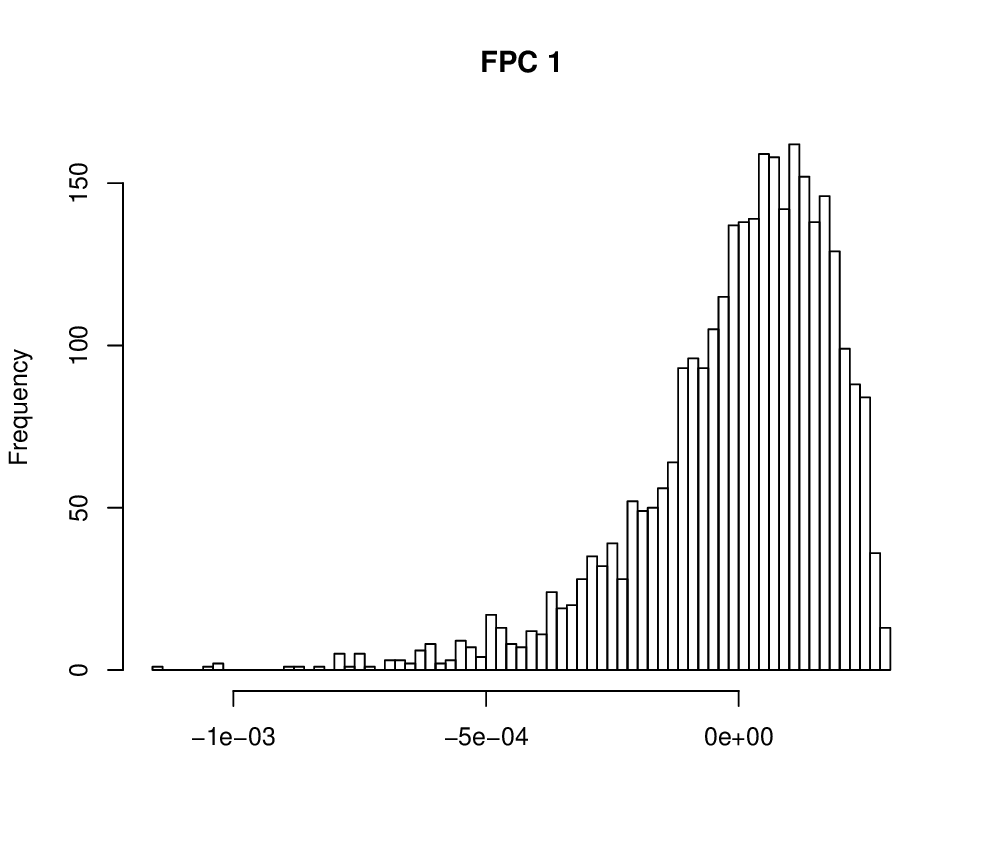} &
\includegraphics[width=.5 \textwidth, height =.3 \textheight]{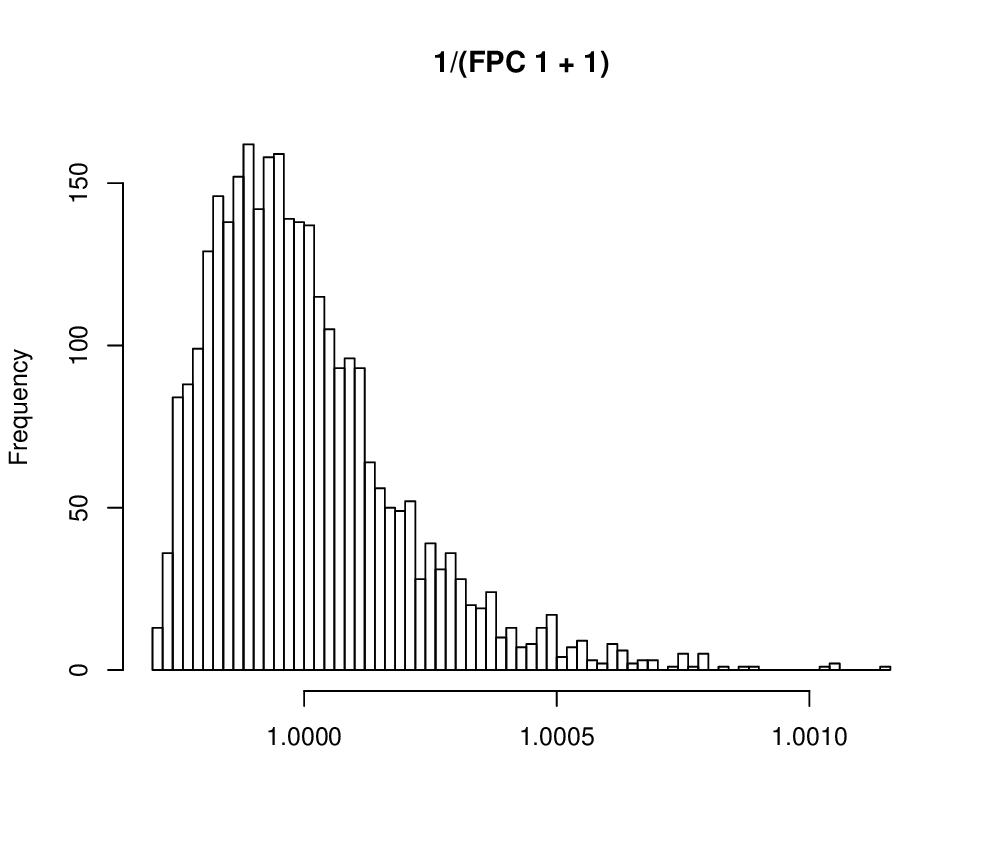} \\
\end{tabular}
\end{center}
\caption{Histogram of absolute frequencies related to the scores of the first functional principal component with and without transformation (right and left panels, respectively).}
\label{fig:histogram}
\end{figure}

In order to fit a probability model to the scores of the first principal component, the histogram of frequencies related to those scores can be seen in Figure \ref{fig:histogram} (left panel). This distribution presents a clear skewness to the left, and then some transformation should be considered in order to fit this data by a known distribution. After some trials, the transformation $1/(\xi_{1}^*+1)$ has been considered. The corresponding histogram of frequencies is shown in Figure \ref{fig:histogram} (right panel).

Taking into account the new transformed data, distributions such as Gamma or Log-normal have been considered without success.

\begin{figure}[h]
\begin{center}
\includegraphics[width=.6 \textwidth, height =.4 \textheight]{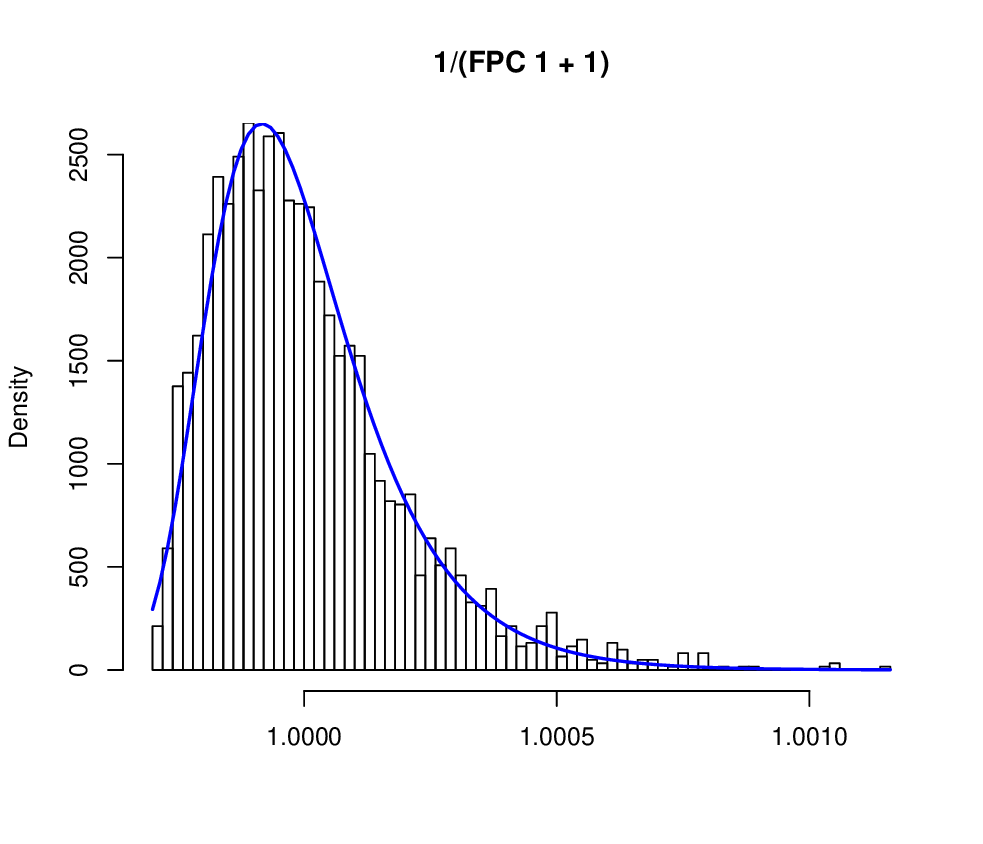} \\
\end{center}
\caption{Density function fitted from a Gumbel distribution with parameters $\mu=0.99992, \; \beta=0.00014,$ with $\mu$ and $\beta$ being
the location and scale parameters, respectively.}
\label{fig:densitygumbel}
\end{figure}

Finally, a Gumbel distribution was considered so that the Kolmogorov-Smirnov goodness-of-fit test provides a
$P-value = 0.06.$ Then, with a significance level of $5\%,$ a Gumbel distribution can be accepted to model the first p.c. scores.
The ML estimation of its parameters is $\mu=0.99992, \; \beta=0.00014,$ with $\mu$ and $\beta$ being the location and scale parameters, respectively.
This fit has been represented in Figure\ref{fig:densitygumbel}.

\section{Conclusions}

Resistive Random Access Memories (RRAMs), based on the resistive switching of transition metal oxide films (TMOs),
is one of the strongest candidates for future nonvolatile applications due to their good scalability, long endurance,
fast switching speed, and ease of integration in the back end of the line of CMOS processing.

This paper introduces a new method to model RRAM devices that can easily account for their current calculation and their statistical
variability description. The methodology is based on a powerful tool of functional data analysis that allows the device current calculation
making use of just one random parameter. A three step procedure based on curve registration, basis representation by P-spline smoothing  and FPCA
orthogonal decomposition, is developed. In this manner, previous complicated compact models for RRAMs can be simplified. In addition, variability, a key
issue to take into consideration prior to industrial use of RRAMs, can be analyzed in an intuitive way, considering a probability function to describe
accurately the distribution of the only parameter employed in the model. This new implementation represents a step forward in the simplicity-accuracy
dilemma that is always presented in the compact modeling context since a reasonable accuracy has been obtained with just one parameter.
Two random parameters can be used if a more restrictive fitting process is needed; nevertheless, two parameters is also a low number compared to
the set of parameters employed in other modeling approaches.
Further research studies focused on the physical meaning of this parameter and other higher
order principal components can be easily performed by employing this new technique.
Other lines of future research will be based on using advanced functional regression models
(functional principal component and partial least squared regression \cite{Aguilera2016}, functional logit regression \cite{AguileraMorillo2012},
functional analysis of variance \cite{Zhang2014}, ...) for modelling important scalar variables related with the design of RRAM devices.

\section*{Acknowledgments}

We thank the Spanish Ministry of Economy and Competitiveness for Projects MTM2017-88708-P  and TEC2014-52152-C3-2-R (also supported by the FEDER program). We would like also to thank F. Campabadal and M. B. Gonz\'alez from the IMB-CNM (CSIC) in Barcelona for fabricating and providing the experimental measurements of the devices employed here.

\section*{References}

\bibliography{Biblio_engineering_revised}

\end{document}